\documentclass[letterpaper,10pt]{article} 

\usepackage{opticameet3} 
\usepackage{subfig} 
\usepackage{multicol}

\newcommand\authormark[1]{\textsuperscript{#1}}

\usepackage{amsmath,amssymb}
\usepackage[colorlinks=true,bookmarks=false,citecolor=blue,urlcolor=blue]{hyperref} 

\begin{document}

\title{Simultaneous Pre-compensation for Bandwidth Limitation and Fiber Dispersion in Cost-Sensitive IM/DD Transmission Systems}

\author{Zhe Zhao,\authormark{1} Aiying Yang,\authormark{1,*} Xiaoqian Huang,\authormark{2} Peng Guo,\authormark{1} Shuhua Zhao,\authormark{1} Tianjia Xu,\authormark{1} Wenkai Wan,\authormark{1} Tianwai Bo,\authormark{1} Zhongwei Tan,\authormark{1} Yi Dong,\authormark{1} and Yaojun Qiao\authormark{2}}

\address{\authormark{1} Key Laboratory of Photonics Information Technology, Ministry of Industry and Information Technology, School of Optics and Photonics, Beijing Institute of Technology, Beijing 100081, China\\
\authormark{2} State Key Laboratory of Information Photonics and Optical Communications, School of Information and Communication Engineering, Beijing University of Posts and Telecommunications, Beijing 100876, China\\}

\email{\authormark{*}yangaiying@bit.edu.cn} 

\begin{abstract}
We propose a pre-compensation scheme for bandwidth limitation and fiber dispersion (pre-BL-EDC) based on the modified Gerchberg-Saxton (GS) algorithm.
Experimental results demonstrate 1.0/1.0/2.0 dB gains compared to modified GS pre-EDC for 20/28/32 Gbit/s bandwidth-limited systems.
\end{abstract}

\section{Introduction}
Driven by applications such as short videos, cloud computing and artificial intelligence, the traffic on data center interconnects has grown exponentially \cite{xie2024multiplication}.
Intensity modulation and direct detection (IM/DD) system has become the mainstream solution for short-reach optical interconnects due to its low cost, low power consumption and simple structure \cite{zhu2022analysis,Zhong:17}. 
Bandwidth limitation (BL) in cost-sensitive IM/DD systems induces inter-symbol interference (ISI) due to high-frequency information loss, while chromatic dispersion (CD) and square-law detection exacerbate signal distortion due to frequency selective fading \cite{wang2023suppressed,Huang:25}.
Therefore, the damage caused by bandwidth limitation and CD remains the main challenge for enhancing C-band transmission performance in IM/DD systems.

Recently, the electronic dispersion pre-compensation (pre-EDC) schemes based on the Gerchberg-Saxton (GS) algorithm have become a research focus. The basic GS pre-EDC considers amplitude at the transmitter and phase at the receiver as degree of freedoms, using the GS iterative algorithm to generate a pre-distorted signal with the ability to combat CD \cite{karar2019iterative}. The modified GS pre-EDC introduces amplitude and phase errors along with related error reversing factors to accelerate convergence and enhance performance \cite{zou2022modified}. 
Moreover, a novel pre-EDC scheme based on polynomial nonlinear filter (PNLF-based pre-EDC) has been proposed \cite{huang2023low}. 
The scheme designs a PNLF to replace the modified GS pre-EDC for signal pre-distortion. 
The feed-forward equalizer (FFE) is needed for post-equalization at the receiver.
In this process, only odd sample information of the signal is used.
To fully utilize the degree of freedoms information in the modified GS pre-EDC, a training-based interleaved dual finite impulse response (FIR) filters-pre-EDC is proposed by designing separate FIR filters for odd and even samples \cite{ni2024odd}. However, few works have investigated the impact of bandwidth limitation on modified GS pre-EDC. 

In this paper, we propose a pre-compensation scheme for both bandwidth limitation and fiber dispersion (pre-BL-EDC) by adding the equalization filter for bandwidth limitation to modified GS pre-EDC. 
Experimental validation on 20/28/32 Gbit/s 50 km on-off keying (OOK) optical fiber transmission systems with a 3-dB bandwidth of 9 GHz demonstrates that pre-BL-EDC without any post-equalization can improve the receiver sensitivity by 1.0/1.0/2.0 dB at the hard-decision forward error correction (HD-FEC) threshold, compared to modified GS pre-EDC with FFE post-equalization.

\section{Principle of Pre-BL-EDC}
\begin{figure}[htbp]
\centering
\includegraphics[width=0.8\linewidth]{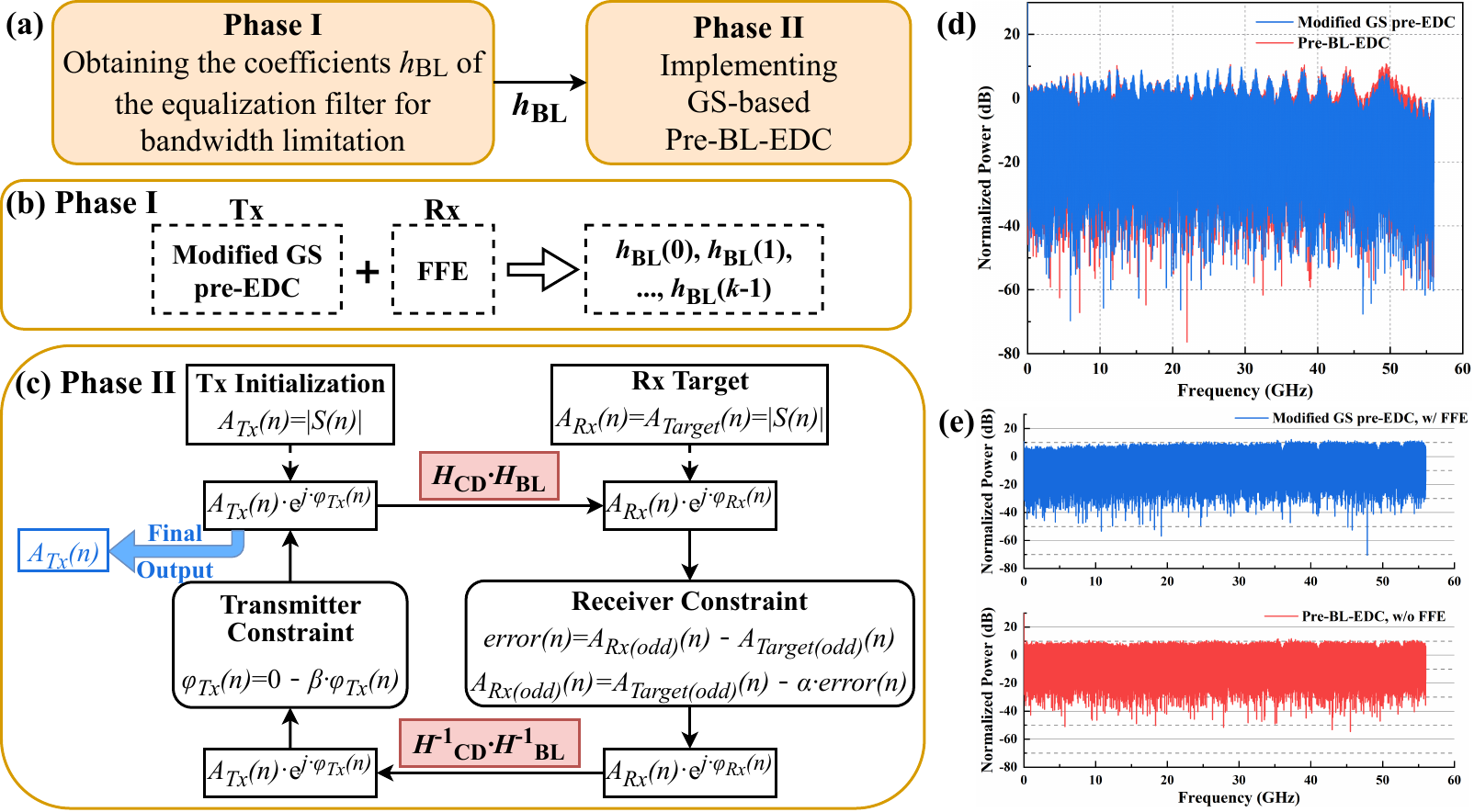}
\caption{(a) Principle diagram of pre-BL-EDC; Specific schematic diagrams of (b) Phase I and (c) Phase II; Frequency spectra of (d) pre-distorted signal and (e) received signal.}
\label{fig_1}
\end{figure}

The schematic of pre-BL-EDC is shown in Fig. \ref{fig_1}(a), consisting of two Phases. 
We design an equalization filter for bandwidth limitation and add it to the modified GS pre-EDC. The coefficients $ h_{\text{BL}} $ of the filter are obtained from Phase I.
Specifically, in Phase I as shown in Fig. \ref{fig_1}(b), modified GS pre-EDC \cite{huang2023low} is applied at the transmitter.
And the receiver uses FFE, whose tap coefficients are then stored as $ h_{\text{BL}} $.
In Phase II as shown in Fig. \ref{fig_1}(c), the Z-domain transfer function of the added filter can be expressed as: $ H_{\text{BL}}(z) = \sum_{n=0}^{k-1} h_{\text{BL}}^{-1}(n) z^{-n} $, where $k$ is the number of taps. 
Then the transfer function $ (H_{\text{CD}} \cdot H_{\text{BL}}) $ is obtained to simulate the fiber CD and system bandwidth limitation, and the rest of the iterative process is identical to that of modified GS pre-EDC. 
As a result, pre-BL-EDC enables the output pre-distorted signal $ A_{Tx}(n) $ to compensate for bandwidth limitation and CD, eliminating the need for FFE post-equalization at the receiver.
Fig. \ref{fig_1}(d) and (e) compare the spectra of the pre-distorted signals and received signals with modified GS pre-EDC and pre-BL-EDC.
Modified GS pre-EDC requires FFE to post-equalize band-limited impairments, which amplifies the high-frequency noise of the received signal. 
Fortunately, pre-BL-EDC can solve this problem. Its pre-distorted signal exhibits slightly higher high-frequency components than modified GS pre-EDC, which can mitigate the impact of system bandwidth limitation, making the received signal spectrum flatter and requiring no post-equalization.

\section{Experimental setup and results}
\begin{figure}[htbp]
\centering
\includegraphics[width=0.8\linewidth]{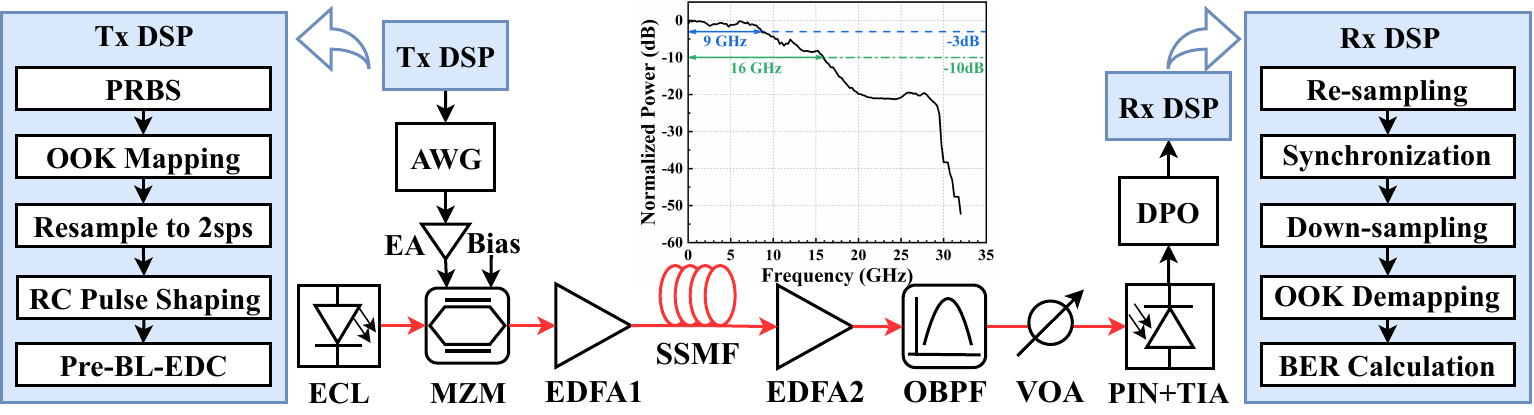}
\caption{Experimental setup and DSP flow for C-band 20/28/32 Gbit/s 50 km OOK IM/DD system with a 3-dB bandwidth of 9 GHz; The inset shows the frequency response of optical BTB channel.}
\label{fig_2}
\end{figure}

We transmit 20/28/32 Gbit/s OOK signals over 50 km standard single-mode fiber (SSMF) in a bandwidth-limited system to evaluate pre-BL-EDC, as shown in Fig. \ref{fig_2}.
At the transmitter, a $ 2^{15} $-length pseudo-random binary sequences (PRBS) is mapped into OOK symbols, upsampled to 2 samples per symbol (sps), and pulse-shaped by a raised cosine (RC) filter with a roll-off factor of 1. Subsequently, pre-BL-EDC is applied to generate a pre-distorted signal resistant to system bandwidth limitation and fiber CD. The signal is converted to analog signal by an 8-bit 65 GSa/s arbitrary waveform generator (AWG, Keysight 8195A) with 20 GHz bandwidth and then amplified by a 19 dB electrical amplifier (EA, SHF 100APP) with 12 GHz bandwidth. The optical carrier is generated by an external cavity laser (ECL, CoBrite DX) with 10 MHz linewidth at 1550 nm. A 40 GHz Mach-Zehnder Modulator (MZM, Fujitsu FTM7937EZ200) is used for electro-optic conversion. Then, the optical signal is amplified to 5 dBm by the first erbium-doped fiber amplifier (EDFA1) before transmission over 50 km SSMF. 
The second EDFA (EDFA2) compensates for the transmission loss, and the amplified spontaneous emission (ASE) noise is filtered by an optical bandpass filter (OBPF, Alnair labs BVF-200CL) with 0.4 nm bandwidth. 
At the receiver, the received optical power (ROP) is varied by a variable optical attenuator (VOA). A 40 GHz PIN with a trans-impedance amplifier (PIN-TIA, MITSUBISHI FU-397SPP) performs photoelectric conversion. 
The electrical signal is captured by an 8-bit 80 GSa/s digital phosphor oscilloscope (DPO, LeCroy 10-36ZiA) with 36 GHz bandwidth, followed by offline digital signal processing (DSP) for resampling, synchronization, downsampling, demapping, and bit error ratio (BER) calculation. 
The inset shows the 3-dB and 10-dB bandwidth of the system are approximately 9 GHz and 16 GHz, respectively, reflecting the insufficient available bandwidth for 20/28/32 Gbit/s OOK signals.

\begin{figure}[htbp]
\centering
\includegraphics[width=0.8\linewidth]{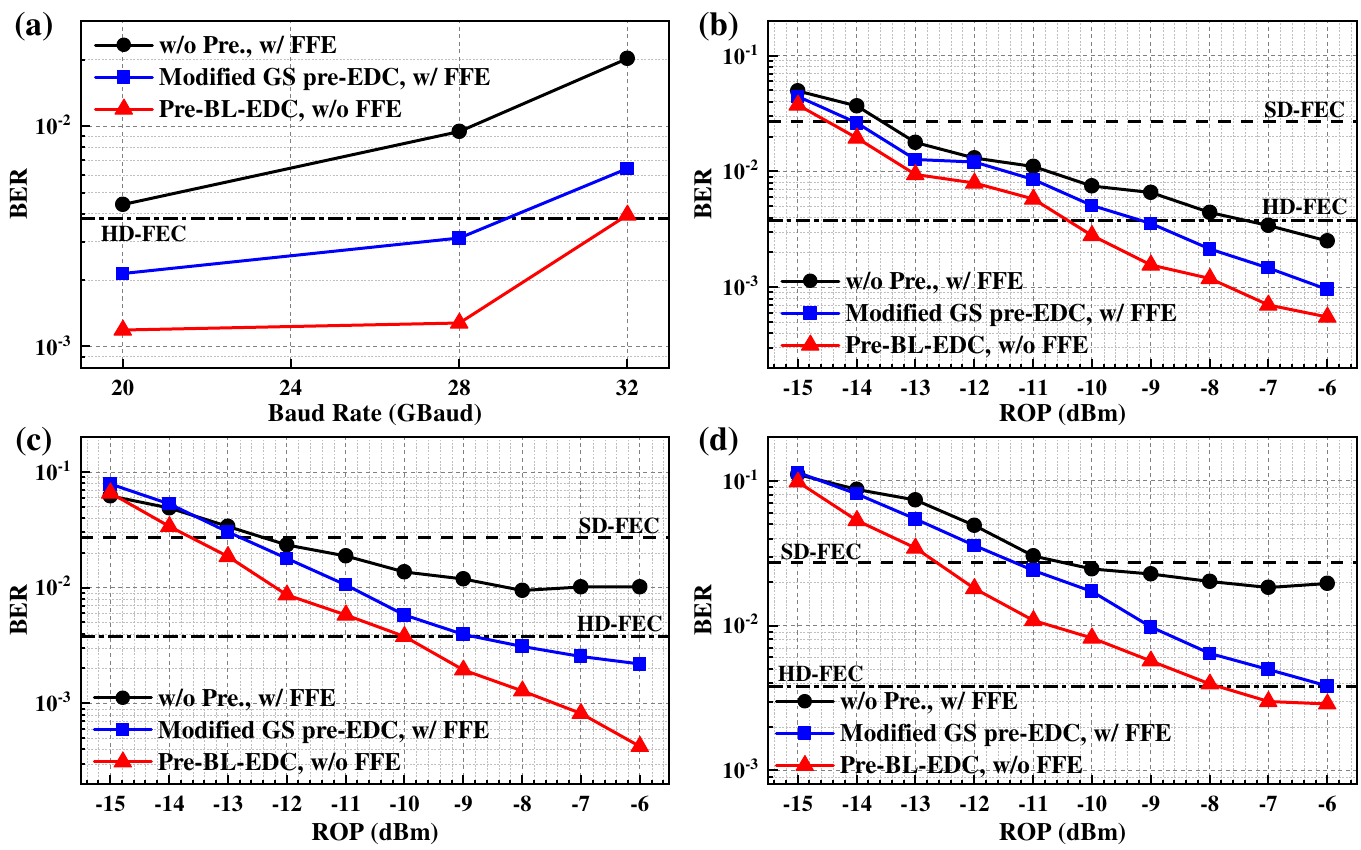}
\caption{Experimental results in C-band 50 km OOK system with a 3-dB bandwidth of 9 GHz: (a) BER versus baud rate at ROP $= -$8.0 dBm; BER versus ROP at (b) 20 Gbit/s, (c) 28 Gbit/s and (d) 32 Gbit/s.}
\label{fig_3}
\end{figure}

In the experiment, the system bandwidth is fixed, so the different bandwidth limitations are realized by changing the baud rate of the OOK signal. 
Fig. \ref{fig_3}(a) compares the BER performance obtained with three schemes: without pre-equalization, modified GS pre-EDC and pre-BL-EDC. 
Among them, without pre-equalization and modified GS pre-EDC require FFE post-equalization, and the FFE taps are optimized based on BER.
Without pre-equalization, the BER of signals at all baud rates fails to reach the HD-FEC threshold.
Modified GS pre-EDC can reduce the BER, but for a 32 Gbit/s signal, the BER cannot reach the HD-FEC threshold due to the more severe bandwidth limitation.
However, pre-BL-EDC demonstrates excellent BER performance, enabling it to satisfy the HD-FEC threshold without post-equalization. This indicates that pre-BL-EDC improves the achievable symbol rate for cost-sensitive IM/DD bandwidth-limited systems.
Fig. \ref{fig_3}(b), (c) and (d) compare the receiver sensitivity of the three schemes at different rates.
For a 20 Gbit/s signal, it can be observed that pre-BL-EDC provides a 1.0 dB receiver sensitivity gain at the HD-FEC threshold compared to modified GS pre-EDC. For a 28 Gbit/s signal, bandwidth limitation becomes severe, whereas pre-BL-EDC still offers a 1.0 dB receiver sensitivity advantage. 
For a 32 Gbit/s signal, modified GS pre-EDC can hardly fall below the HD-FEC threshold, whereas pre-BL-EDC reaches the HD-FEC at ROP of -8.0 dBm, providing a 2.0 dB receiver sensitivity gain.

\section{Conclusion}
In this paper, a pre-BL-EDC scheme is proposed to jointly compensate for system bandwidth limitation and fiber CD. 
The proposed scheme integrates an equalization filter for bandwidth limitation into the modified GS pre-EDC.
Experimental results on C-band 50 km IM/DD optical fiber transmission systems demonstrate that pre-BL-EDC exhibits superior bandwidth limitation tolerance over modified GS pre-EDC, without any post-equalization.
With the system 3-dB bandwidth of 9 GHz, the receiver sensitivity of 20/28/32 Gbit/s OOK signals using pre-BL-EDC is improved by 1.0/1.0/2.0 dB over modified GS pre-EDC at the HD-FEC threshold, respectively.
The results envision that the proposed pre-BL-EDC scheme provides a new hardware-friendly pre-compensation paradigm for cost-sensitive IM/DD optical transmission systems.

\end{document}